\documentclass{elektr}
\usepackage{hyperref}
\hypersetup{
colorlinks=true,
urlcolor=blue,
citecolor=blue}
\usepackage[all]{xy,xypic}
\usepackage{amsfonts,amssymb,amsmath,amsgen,amsopn,amsbsy,theorem,graphicx,epsfig}
\usepackage{eufrak,amscd,bezier,latexsym,mathrsfs,eurosym,enumerate}
\usepackage[utf8]{inputenc}
\usepackage[english]{babel}
\usepackage{cleveref, multirow}
\usepackage[dvipsnames]{xcolor}
\usepackage[pagewise]{lineno}

\yil{}
\vol{}
\fpage{}
\lpage{}
\doi{}

\title{A fast text similarity measure for large document collections using multi-reference Cosine and genetic algorithm}

\author[H. Mohammadi and S. H. Khasteh]{
\textbf{Hamid Mohammadi$^{1}$\thanks{mohammadi2823@email.kntu.ac.ir}~, Seyed Hossein Khasteh$^{1}$}\\
$^{1}$Department of Computer Engineering, K. N. Toosi University of Technology, Tehran, Iran
\\ [1.8em]

\rec{.201}
\acc{.201}
\finv{..201}
}

\def\E{\ifmmode{\mathbb E}\else{$\mathbb E$}\fi} 
\def\N{\ifmmode{\mathbb N}\else{$\mathbb N$}\fi} 
\def\R{\ifmmode{\mathbb R}\else{$\mathbb R$}\fi} 
\def\Q{\ifmmode{\mathbb Q}\else{$\mathbb Q$}\fi} 
\def\C{\ifmmode{\mathbb C}\else{$\mathbb C$}\fi} 
\def\H{\ifmmode{\mathbb H}\else{$\mathbb H$}\fi} 
\def\Z{\ifmmode{\mathbb Z}\else{$\mathbb Z$}\fi} 
\def\P{\ifmmode{\mathbb P}\else{$\mathbb P$}\fi} 
\def\T{\ifmmode{\mathbb T}\else{$\mathbb T$}\fi} 
\def\SS{\ifmmode{\mathbb S}\else{$\mathbb S$}\fi} 
\def\DD{\ifmmode{\mathbb D}\else{$\mathbb D$}\fi} 

\newcommand{\bse}{\begin{subequations}}
\newcommand{\ese}{\end{subequations}}
\newcommand{\ben}{\begin{enumerate}}
\newcommand{\een}{\end{enumerate}}
\newcommand{\bens}{\begin{enumerate*}}
\newcommand{\eens}{\end{enumerate*}}
\newcommand{\be}{\begin{equation}}
\newcommand{\ee}{\end{equation}}
\newcommand{\bea}{\begin{eqnarray}}
\newcommand{\eea}{\end{eqnarray}}
\newcommand{\baa}{\begin{eqnarray*}}
\newcommand{\eaa}{\end{eqnarray*}}
\newcommand{\bc}{\begin{center}}
\newcommand{\ec}{\end{center}}

\newcommand{\vs}{\vspace}

\theoremstyle{corollary}

\theoremstyle{lemma}

\theoremstyle{proposition}

\theoremstyle{axiom}

\theoremstyle{conjecture}

\theoremstyle{example}

\theoremstyle{definition}

\theoremstyle{remark}

\begin{document}

\maketitle

\begin{abstract}One of the critical factors that make a search engine fast and accurate is a concise and duplicate free index. In order to remove duplicate and near-duplicate (DND) documents from the index, a search engine needs a swift and reliable DND text document detection system. Traditional approaches to this problem, such as brute force comparisons or simple hash-based algorithms are not suitable as they are not scalable and are not capable of detecting near-duplicate documents effectively. \\
In this paper, a new signature-based approach to text similarity detection is introduced, which is fast, scalable, reliable, and needs less storage space. The proposed method is examined on standard text document datasets such as CiteseerX, Enron, Gold Set of Near-duplicate News Articles, and other similar datasets. The results are promising and comparable with the best cutting-edge algorithms, considering accuracy and performance. The proposed method is based on the idea of using reference texts to generate signatures for text documents. The novelty of this paper is the use of genetic algorithms to generate better reference texts.

\end{abstract}

\section{Introduction}
\label{intro}
The main task of a search engine is searching! Therefore, it has to acquire a set of web pages to search through them, which is called index. A search engine works fast and reliable when its index is as concise as possible without missing any possible web pages. For this purpose, the DND documents must be removed from the index. Another component of a typical search engine which is prone to DND text documents problem is a crawler. The crawler is a component in a search engine which has the responsibility of surfing the web and downloading web pages to index. A crawler faces the problem of DND web pages since a significant number of web pages on the web are DND web pages \cite{broder1997syntactic} \cite{fetterly2003evolution} \cite{henzinger2006finding}. In order to have a formal definition of DND web pages, we can say that duplicate web pages are the pages that are identical in terms of content, but they are accessible using multiple URLs. On the other hand, near-duplicate pages are the web pages with slight differences, such as changed date or some other minor edits, but they are not the same \cite{manku2007detecting} \cite{xiao2011efficient}. One of the problems with the DND web pages is that they increase the size of the search index. Such an index could reduce the quality of search engine results and increase the computational power needed to perform all kinds of tasks on them \cite{fetterly2003evolution}. An index works better when the search engine is able to detect the duplication or near-duplication web pages in a speedy and accurate manner \cite{fetterly2003evolution} \cite{henzinger2006finding} \cite{conrad2003online}. The common approaches to overcome this problem include the brute force approaches and hash-based algorithms. Approaches using brute force techniques compare each document with all other documents using a similarity measure like Cosine \cite{broder1997syntactic} or Jaccard \cite{theobald2008spotsigs}. These approaches are not suitable for large-scale document collections, containing hundreds of thousands of text documents such as the World Wide Web. These approaches are not efficient since all documents must be compared one by one. The time complexity of comparing all documents is $O \left(n^2\right) * L$ where L is the mean length of documents. One approach to solve this problem is to reduce the L using hashing methods. These methods generate a signature (which is much smaller than the document itself) for each document. For example, a 128 or 160-bit length hash is created for each document using MD5 or SHA1 algorithms. As a result, it is possible to compare the signature of documents instead of comparing the documents and reduce the processing time. In these methods, the required computational power is decreased drastically. The most significant shortcoming of such methods is that most of these approaches are incapable of detecting near-duplication as the amount of similarity between two documents cannot be measured using hashes generated by these hashing methods \cite{manku2007detecting}. \\
In this study, a new approach to overcome these odds is proposed. The new approach is based on the multi-reference cosine text similarity algorithm \cite{multireferencecosine}. The new approach is scalable, reliable, and fast. The proposed method can also be used in some other applications such as document clustering, plagiarism detection, and recommenders systems. The main contribution of the proposed method is the achievement of the highest recall in near-duplicate detection task among state of the art approaches to near-duplicate document detection algorithms in large document collections. \\
The rest of this paper is structured as follows: In section \ref{sec:1}, some of the related works are reviewed. In section \ref{sec:2} the new approach is explained. Section \ref{sec:4} will show the experimental results. Finally, Section \ref{sec:5} presents conclusions and future works.

\section{Related Works}
\label{sec:1}
Search engines are the gateway to the web \cite{ntoulas2004s}. Indexes are one of the key components of a search engine and have a deep influence on its performance \cite{croft2010search}. One of the properties of a good index is that it is not suffering from DND entries. \\
Several studies have been conducted in order to solve the problem of finding and removing the DND documents and web pages. These studies could be classified into two major classes. URL based methods and content-based methods. The URL based methods find common patterns in the URLs of web pages and recognize DND web pages using these patterns \cite{koppula2010learning}. One of the first works on this branch of approaches was done by Bar-Yossef et al. \cite{bar2009not}. Later Dasgupta et al. \cite{dasgupta2008duping} and Koppula et al. \cite{koppula2010learning} extended his works on detecting DND web pages using URLs. Koppula analyzed web servers log to find patterns in URLs that are pointing to DND web pages. \\
The other class of DND text detection methods is content-based methods class. The new content based attempts to solve the problem of DND text documents detection were brute force approaches. In these approaches, a server would compare each document with all other documents using a similarity measure such as cosine text similarity \cite{shivakumar1995scam}. Brute force methods are accurate because they compare all the documents one by one. However, the disadvantage of these approaches is their performance. Therefore, these approaches are not suitable for applications with a great number of text documents. Later, researchers focused on finding DND documents with a lesser amount of computational power and storage space. The cornerstone of these studies was the works of Manber \cite{manber1994finding} and Heintze \cite{heintze1996scalable}, which were focused on the adjacent characters' resemblance. Meanwhile, Brin \cite{brin1995copy} came up with a system that used hashes to detect copyright violations. Later, Broder et al. \cite{broder1997syntactic} introduced the shingling algorithm, which was used in the AltaVista search engine as a duplicate document detection algorithm. Border's algorithm does not use any prior knowledge of the target language, which makes it more suitable for the multi-lingual ecosystem of the web. The Border's method uses the Jaccard similarity measure to compare documents shingles. Next year, Locality Sensitive Hashing or LSH, which is an approximate approach to find DND documents, introduced by Indyk et al. \cite{indyk1998approximate}. The approximate approach to DND text detection ensures the performance of the algorithm but increases the number of false-positive and false-negative decisions \cite{schleimer2003winnowing}. I-match \cite{chowdhury2002collection} is another innovation to find DND text documents, which uses lexical methods. I-match uses a lexicon which is created from a large text corpus and creates a signature for each document, using the SHA1 hashing algorithm. The similarity of the signatures generated by the I-match algorithm will show the probability of the duplication. Later, Sarawagi et al. \cite{sarawagi2004efficient} used the inverted-index method, which is only able to find duplicate documents. \\
One of the most efficient and frequently-used algorithms for DND document detection is introduced by Charikar \cite{charikar2002similarity}. Charikar's algorithm, Simhash, uses dimensionality reduction techniques to generate fixed-length hashes for each document. The Simhash algorithm has shown promising results \cite{pamulaparty2013novel} and is being used by many search engines around the world, such as Google \cite{manku2007detecting}. Henzinger \cite{henzinger2006finding} combined Broder's and Charikar's algorithms to achieve higher precisions. \\
Another state of the art algorithm is SpotSigs \cite{theobald2008spotsigs}. SpotSigs showed that some parts of documents have a higher impact on the measured similarity of the documents, comparing to the other parts. Hajishirzi et al. \cite{hajishirzi2010adaptive} introduced a domain-specific algorithm which uses a real-valued k-gram summary vector as a signature for each document and can be adapted to use different similarity measures like Cosine and Jaccard to detect duplicate and near-duplicate text documents. \newline
As the amount of available data and the number of web pages are increasing rapidly in the recent years, the use of big data (map-reduce) techniques is getting common among the studies for DND document detection. Lin \cite{lin2009brute} and Vernica's \cite{vernica2010efficient} works are of this type, and they tried to adapt current methods to big data applications and frameworks. \\
Some of the recent studies are focused on DND document detection systems or new hybrid approaches which combine the previous algorithms and enhances the precision or performance. For example, the works of Pamulaparty et al. \cite{pamulaparty2014near} can be mentioned. In his research, a new architecture for DND document detection is introduced. The system introduced by Pamulaparty first parses a web page and extracts its texts from it. Then, after stop word removing and stemming, it adds the new document to its database and checks the database for documents with the highest common words and marks them as DND texts. Varol et al. \cite{varol2015detecting} introduced a new hybrid approach to DND document detection. Varol combined shingling with Jaro distance and word usage frequency to enhance the shingling algorithm \cite{varol2015detecting}. \\
Zhang et al. \cite{zhang2016effective} introduced one of the latest and most efficient methods to find DND text documents. Zhang used the idea of Normalized Compression Distance, or NCD \cite{cilibrasi2005clustering} and combined it with the idea of hashing in order to solve the problem of NCD with medium and large files. NCD is a similarity metric that is universal and parameter-free. NCD is based on the Kolmogorov complexity \cite{li2008p}. The main idea of NCD is that if two text documents have more common information, the result of compressing these two files together is smaller. In other words, the shorter the compressed form of two text documents is, the more similar those two texts are. The Zhang method, which is called SigNCD is both accurate and high-performance. Zhang also introduced another algorithm which is called SpotSigNCD \cite{zhang2016effective}. The SpotSigNCD is based on the SigNCD but uses the stop word-spot signature extraction method instead of the original signature extraction method of the SpotSig \cite{zhang2016effective}. SpotSigNCD has better precision than the SigNCD, but its recall is not as good as the SigNCD \cite{zhang2016effective}. In this paper, a new signature-based method is introduced, which has competitive accuracy and performance.

\section{Proposed algorithm}
\label{sec:2}
\begin{figure}[h!]
\begin{center}
\includegraphics[width=9.0cm]{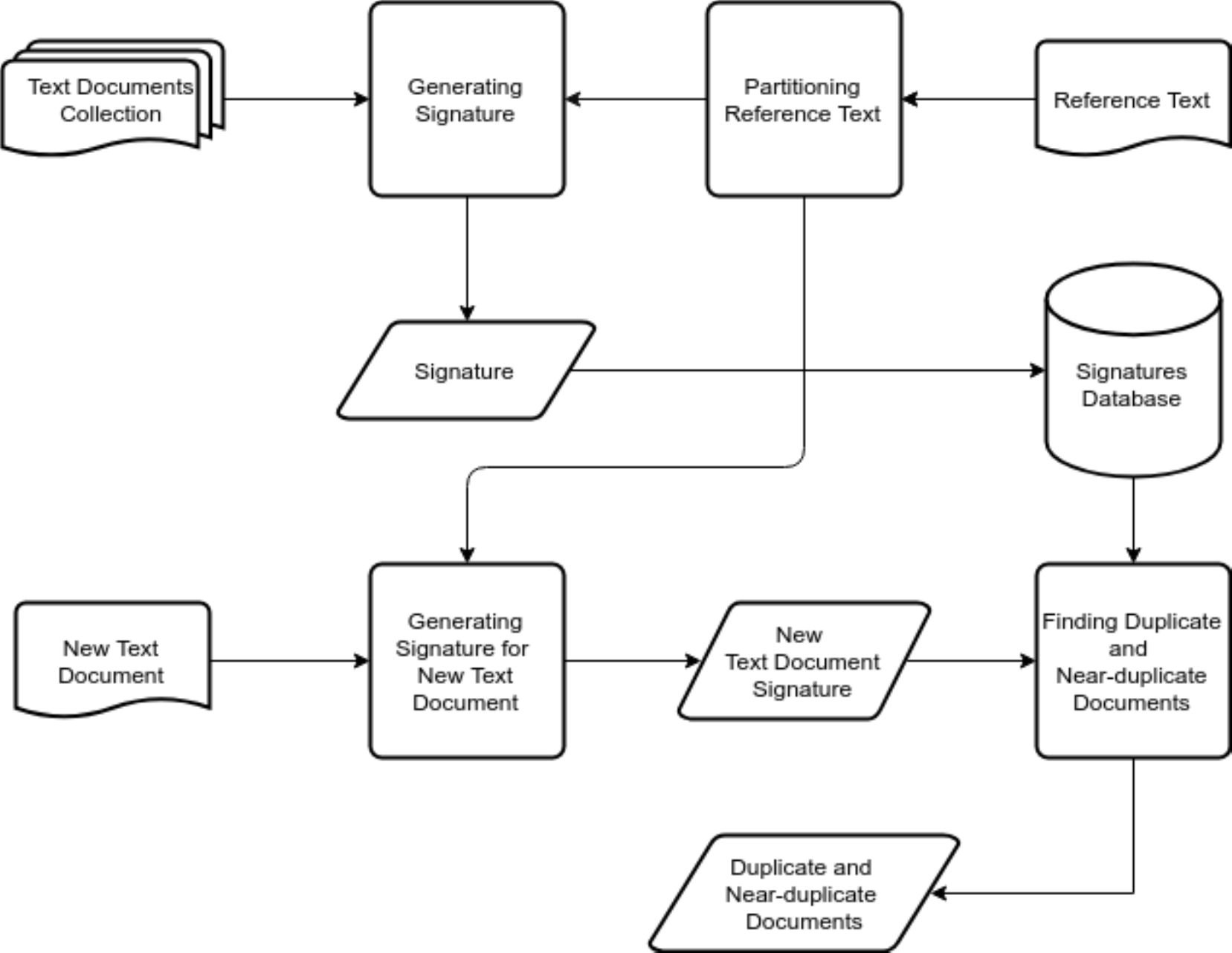}
\caption{Structure of DND documents detection system using multi-reference Cosine.}
\label{fig:1}
\end{center}\vs{-4mm}
\end{figure}

The proposed method works based on an idea called multi-reference cosine text similarity algorithm \cite{multireferencecosine}. Each text (reference text and text document) is considered as a sequence of 3-grams. In this algorithm, in order to generate the signature of the text document ($D_i$), it compares ($D_i$) with different parts of reference text using cosine text similarity measure. Cosine text similarity algorithm is chosen as our core similarity measurement algorithm as the Cosine text similarity algorithm is more accurate than other similar algorithms such as Pearson Correlation Coefficient (PCC), Jaccard, and Mean Square Difference (MSD) in terms of measuring the similarity or difference between texts \cite{liu2014new}. The result is a decimal number for each comparison. The algorithm puts these decimal numbers altogether to create a vector and considers this vector as the signature of ($D_i$). The algorithm uses the signature to calculate the similarity between different text documents. The generated signature has the property that if the original text documents are similar, the signatures of those text documents are similar as well. Therefore the algorithm can detect DND text documents using this signature. The proposed method could be used for measuring the similarity degree between documents in high volume text document datasets in a fast and efficient way. The overall structure of a system, using the proposed algorithm to find DND documents, is shown in Figure \ref{fig:1}.

The system in Figure \ref{fig:1} consists of text documents, reference text, document-reference comparison, signatures database, and similarity measurement. The system components are explained as follows:

\subsection{Text Documents}
\label{sec:3.1}
The text documents could be text files, web pages, or any other type of files which consist of text. Text documents can be seen in two parts of Figure \ref{fig:1}.  \\
``Text Documents Collection'': The Text Documents Collection is a database which contains all previously collected text documents. \\
``New Text Document'': The new text document is a text document which the system wants to find its DND text document in the database.

\subsection{Reference Text}
\label{sec:3.2}
The reference text is one of the main components of the Multi-reference cosine algorithm. The reference text is a sequence of 3-grams, which is used by the algorithm to generate a signature for each text document. N-grams are the essential elements of comparison in cosine text similarity algorithm, which is the core of multi-reference cosine algorithm. Longer N-grams can be more effective in terms of measuring the similarity or difference between two texts. On the other hand, a greater value of N means a higher number of possible permutations for N-gram, which makes the reference text search space larger, hence increasing the processing power and time required to find a desirable reference text. Considering this trade-off between shorter and longer N-grams, 3-grams are found optimum in terms of effectiveness in measuring the similarity between documents and performance \cite{multireferencecosine}. Reference text could be generated with the help of methods such as information gain method. In this paper, a new method of generating the reference text is proposed. The new method uses genetic algorithms to find better reference text to achieve more accuracy and performance for the multi-reference cosine text similarity algorithm. By using this new method of reference text generation, the performance of the whole system outperforms some of the best state of the art algorithms like Simhash \cite{charikar2002similarity} which Google reported being using it as DND web page detection algorithm in its search engine \cite{manku2007detecting}. Using this new method, the previous method of generating reference text that uses information gain theory to minimize the Mean Absolute Error of the multi-reference cosine algorithm \cite{multireferencecosine} is outperformed too.

\subsection{Reference Text partitioning}
\label{sec:3.3}
To generate a signature, multi-reference cosine text similarity algorithm splits the reference text into several parts. It creates multiple reference texts from the original reference text. The Multi-reference cosine algorithm compares a document with the reference text parts using cosine text similarity measure. The result of each comparison is a decimal value between zero and one. The greater the number is, the more similar is the text document to that part of the reference text. In other words, the algorithm examines how frequent the 3-grams of that specific reference text are in the selected text document. The best 3-grams to be a part of the reference text, are the 3-grams that have the highest information about how similar or different that two documents are. Therefore, the existence or absence of each 3-gram inside the reference text may have a considerable impact on the signature of a document. Given this fact, if the algorithm compares a document with one reference text that contains many 3-grams, it has missed some of the information that each 3-gram could have given us. For example, there may be a reference text that contains ($3-gram_{n}$) and ($3-gram_{m}$). \\
On the other hand, there is a document ($D_i$) which contains ($3-gram_{n}$) and have high similarity with the part of reference text that containing ($3-gram_{n}$) and zero similarity to the other parts of reference text. There is also another document ($D_j$) which contains ($3-gram_{m}$) and have high similarity with the part of reference text containing the ($3-gram_{m}$) and no similarity with the other parts. These two documents get the same signatures because they contain the same number of common 3-grams with the reference text, and they will be detected similar. While, if the algorithm divides the reference text into several parts, each containing a limited number of 3-grams, it can harness more information from what that each 3-gram can give us about a document. If the reference text contains all possible 3-grams in the character set that the documents are made of, by reducing the size of each partition of reference text to a single 3-gram, the algorithm is creating the vector space model of the document, which is somehow a complete numeric representation of each document. In the scenario above, having the exact vector space model of each document will guarantee the cosine text similarity accuracy. However, its performance is similar to a naive brute force text comparison algorithm, which is not desirable for large text document collections. \\
Nevertheless, all the 3-grams are not important equally regarding similarity detection, and some 3-grams contain more information than the others. By selecting these important 3-grams and removing other 3-grams from reference text, the multi-reference cosine text similarity algorithm can reduce reference text's size drastically and increase the performance. The exact number of partitions depends on the desired accuracy and performance. The effect of the number of partitions on the final accuracy and performance will be discussed in section \ref{sec:4}.

\subsection{Signature Generation}
\label{sec:3.4}
As mentioned before, in order to generate a signature for each text document, the algorithm compares the text document with each reference text part using cosine text similarity measure. The result of each comparison is a decimal value between zero and one. The algorithm stores the comparison results in an array. It uses this array as the signature of the text document.

\subsection{Finding DND Documents}
\label{sec:3.5}
Since the signature is an array of decimal values, it can be considered as a vector. In order to compare documents in this system, the algorithm can compare their signatures, which are stored in a database, using a vector similarity measure such as cosine text similarity. To find DND text documents, the algorithm can mark text documents with a very high (higher than or equal to a predefined threshold $t_{1}$) degree of similarity as candidate duplicate documents and mark documents between two predefined thresholds $t_{1}$ and $t_{2}$ (higher than or equal to $t_{2}$ and less than $t_{1}$) as candidates for near-duplicate documents. $t_{1}$ and $t_{2}$ thresholds take values between 0 and 1, which are determined empirically and using trial and error for each specific application. A multi-reference cosine algorithm with greater values for $t_{1}$ and $t_{2}$ finds documents with a higher degree of similarity to the original document as DND documents, in comparison to the same algorithm with lower values for $t_{1}$ and $t_{2}$.

\subsection{Generating Reference Text}
\label{sec:3.6}
The reference text has a significant impact on the performance of the newly proposed method. Therefore, choosing an excellent reference text is one of the most critical parts of the multi-reference cosine algorithm. Before this study \cite{multireferencecosine}, the idea of information gain is used to generate reference texts. The results were acceptable, yet a better reference text is needed to acquire better results. In this study, a genetic algorithm is used in order to generate a better reference text. The reference text is a sequence of characters so that it can be considered as a chromosome, and accordingly, it is possible to use genetic algorithms to find good reference texts. The flowchart of used genetic algorithm is presented in Figure \ref{fig:2}. The components of such a system are explained as follows:

\begin{figure}[h!]
\begin{center}
\includegraphics[width=9.0cm]{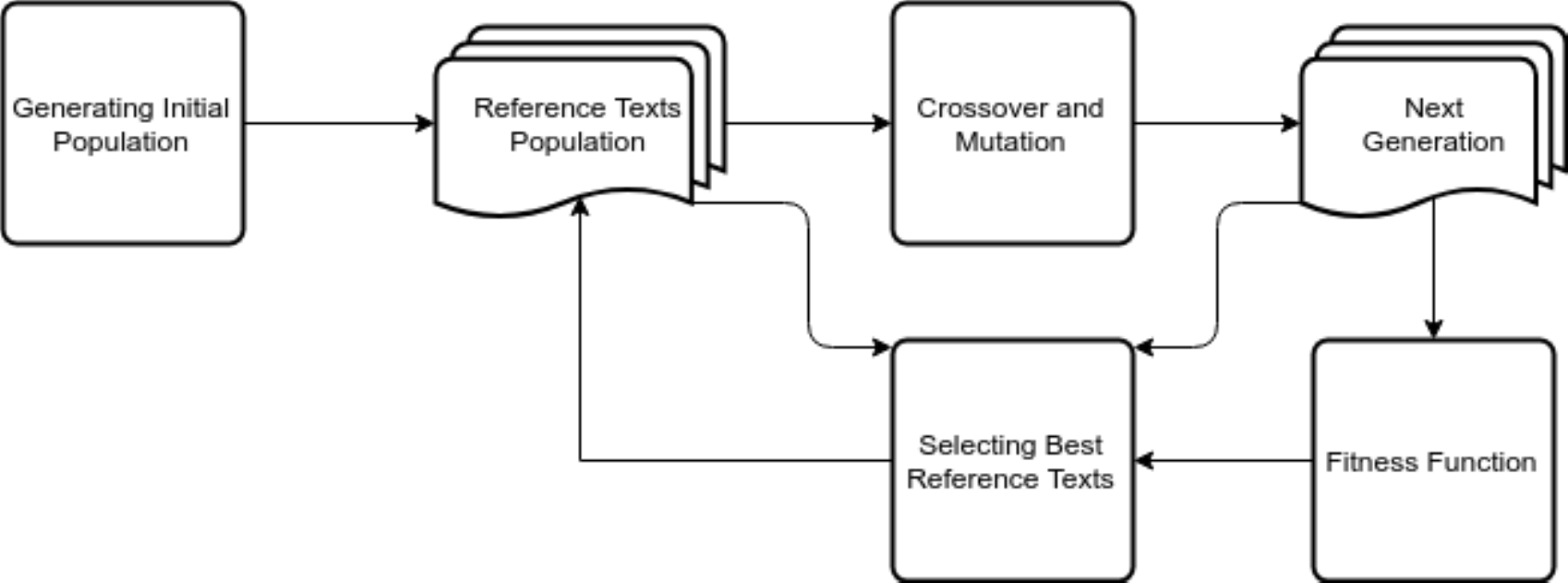}
\caption{Generating reference texts using genetic algorithm.}
\label{fig:2}
\end{center}\vs{-4mm}
\end{figure}

\subsubsection{Initial Population}
\label{sec:3.6.1}
The initial population can be generated using different methods, but in order to increase the quality of the final results and reducing the number of generations needed to reach a desirable result, N-grams with high information gain to generate the reference texts are used. N-gram is an N character long string. In this implementation of the cosine text similarity algorithm, 3-gram model is used in order to compare two texts. 9000 3-grams with highest Tf-idf score to create the initial population of reference texts are extracted from Open American National Corpus \cite{ide2009american}. To score a document using Tf-idf method, the term (in this case 3-gram) frequency in each document and the inverted document frequency (the frequency of documents which the term appeared in them) are used. By using Tf-idf scoring method, we ensure that the selected 3-grams are the 3-grams with highest information gain, which are the most useful 3-grams to determine how much two text documents are similar or different. After selecting 3-grams, the initial population is generated by randomly combining some of the highly scored 3-grams to a fixed length. \\
The final accuracy and performance of the multi-reference cosine text similarity algorithm directly depend on the reference text length. If the algorithm uses a concise reference text, although the reference text is made by using the best possible 3-grams, the algorithm cannot exceed an accuracy limit, because of the limited number of possible 3-grams in a short reference text. \\
On the other hand, if the reference text contains all possible 3-grams in a language or character set, the increase in the reference text length will no longer increase the final accuracy. Besides, the longer reference text makes the multi-reference cosine text similarity algorithm slower. Therefore, in order to achieve good overall accuracy and performance, we should consider choosing an optimized reference text length. In general, there is a tradeoff between accuracy and performance of multi-reference cosine text similarity algorithm. The effect of the different reference text lengths on the final accuracy will be discussed in section \ref{sec:4}.

\subsubsection{Next Generation}
\label{sec:3.6.2}
The genetic algorithm uses methods such as crossover and mutation to generate the intermediate population from the current population. Crossover and mutation can be done in different ways and with different rates. As a result of conducted tests, the best methods and rates that will help to reach the best results in the case of optimizing reference texts are found. The method that is used for the crossover is single cut crossover which means that the algorithm cuts two reference texts in a random length and displaces the cuts. The method used for mutation is to randomly replace 10 percent of 3-grams of a reference text. As we mentioned earlier, 3-grams with higher Tf-idf score make reference texts more effective in terms of accurately measuring the similarity or dissimilarity between two documents; accordingly, the selected 10 percent of 3-grams are replaced with randomly chosen 3-grams from a list of 3-grams with high Tf-Idf score in OANC corpus.

\subsubsection{Fitness Function}
\label{sec:3.6.3}
In this step, as a fitness function, reference texts get scores according to how accurate the multi-reference cosine algorithm can measure the similarity degree between different documents. As mentioned in section \ref{sec:2}, the higher accuracy of Cosine text similarity algorithm compared to other text similarity measures such as Pearson Correlation Coefficient (PCC), Jaccard, and Mean Square Difference (MSD), in terms of calculating the degree of similarity between two texts makes it a gold standard for text similarity \cite{liu2014new}. A version of multi-reference cosine text similarity algorithm is implemented, that is using the generated reference text, over a number of randomly selected documents in a dataset, comparing them one by one. Similar to any form of statistical modeling, the randomness of sampling can affect the outcome of an algorithm, yet, large samples from the original data could provide a precise model. On the other hand, larger samples demand more processing power to process. In this study, sample sizes, which is mentioned for each dataset in section \ref{sec:4.1}, is determined to obtain desired accuracies with the least processing power required. The final score is the mean absolute error of the results of the multi-reference cosine algorithm in comparison with the results of a traditional cosine text similarity algorithm. The fitness function is executed for each reference text in the intermediate population. The final score is shown in Eq. \ref{eq:1}:

\begin{equation}
\resizebox{0.8\textwidth}{!}
{
$Final\ Score = \frac{\sum_{i=0}^{N-1}\sum_{j=i+1}^{N}\mid{}Multi\:reference\:cosine(D_{i}-D_{j})-cosine(D_{i}-D_{j})\mid{}}{\frac{N(N-1)}{2}}$
\label{eq:1}
}   Similar
\end{equation}

Where N is the number of randomly selected documents, $D_{i}$ and $D_{j}$ are $ i^{th}$ and $ j^{th}$ documents in the randomly selected document set and $\frac{N(N-1)}{2}$ is the total number of comparisons.

\subsubsection{Selecting Reference Texts}
\label{sec:3.6.4}
Next, the next generation is selected from the intermediate population. The selection criteria is the score which each reference text acquires through the fitness function. The reference texts with higher scores are selected. The number of selected reference texts is the same as the number of reference texts in the main population. This new population takes the place of the previous main population. \\
The process of generating intermediate population, scoring intermediate population reference texts and selecting the next generation reference texts continues until the accuracy of the whole system using the latest generation of reference texts reaches a predefined threshold. Usually, the average number of needed generations to reach the desired threshold is about 50 generations.

\section{Test Results}
\label{sec:4}
In order to test the newly generated reference texts, the whole system must be tested using the new reference text. The multi-reference cosine is tested using generated reference texts on several datasets. Reported runtimes are computed using a personal computer running Ubuntu 17.04 and OpenJDK Java 1.8.0, with a two core, core i7 CPU and 8 Gigabytes of RAM. Datasets and test results are as follows:

\subsection{Datasets}
\label{sec:4.1}
Several standard datasets that have been used in similar studies are gathered. The datasets are:

\subsubsection{Citeseerx Dataset}
\label{sec:4.1.1}
Citeseerx, is a scientific literature digital library and search engine that is focused on scientific literature in the field of computer and information science. Citeseerx dataset \cite{Citeseerx} contains more than 6 million scientific papers and their metadata. Ten thousand of randomly selected documents in the Citeseerx dataset have been used for the test purposes.

\subsubsection{TREC 2005}
\label{sec:4.1.2}
TREC stands for Text Retrieval Conference. It is a conference that is held annually, and it is focused on encouraging studies in the field of information retrieval. TREC 2005 public spam corpus \cite{TREC} contains 92,189 email messages. Each message is stored in a text file, so it can be used for evaluating the DND text document detection method. The portion of the dataset that is used for the test purposes is consist of ten thousand randomly selected email messages.

\subsubsection{DMOZ}
\label{sec:4.1.3}
DMOZ is an open-content web directory of World Wide Web URLs, which is also known as "open directory" project. The original DMOZ dataset \cite{DMOZ} contains over two million records. About ten thousands of URLs are randomly selected, and the corresponding web pages are downloaded. All HTML tags are removed from the downloaded web pages, and the content of each web page is extracted. The final documents are used for the tests.

\subsubsection{Newsgroups}
\label{sec:4.1.4}
20 newsgroup dataset \cite{Newsgroups} is a collection of about 20,000 newsgroup documents. Ken Lang originally collected this dataset for his ``Newsweeder: Learning to Filter Netnews'' paper \cite{lang1995newsweeder}. The data is categorized into 20 topics. Just as other datasets, ten thousands of documents in this dataset are randomly chosen and are used for the test purposes.

\subsubsection{OpenDNS Public Domains}
\label{sec:4.1.5}
OpenDNS public domains dataset \cite{OpenDNS} is originally a list of several randomly selected domains that are stored on OpenDNS domain name servers. This dataset contains a list of ten thousand domain names. The coresponding web pages of these domain names are downloaded, and all HTML tags are removed from them. The remaining content is used for test purposes.

\subsubsection{Enron}
\label{sec:4.1.6}
Enron was one of the biggest companies in the United States of America in the 90s. Later, in 2001, the company declared bankruptcy. The company faced a financial crisis because of creative and systematic accounting fraud. Enron dataset \cite{Enron} is a large collection of emails from or to the Enron employees. It contains over one hundred thousand files. About ten thousand randomly selected files in this dataset are used for the test purposes.

\subsubsection{Gold Set of Near Duplicate News Articles}
\label{sec:4.1.7}
Gold Set of near Duplicate News Articles \cite{GoldSet} is a dataset consist of over 2,000 news articles clustered into 68 categories. This dataset is originally a part of Stanford Web Base crawl \cite{cho2006stanford}. It is a great dataset to evaluate a near-duplicate detection algorithm because human assessors have manually marked the near-duplicate documents in this dataset. Also, it was used by many researchers studying detection of DND text documents in large collections, to evaluate their algorithms \cite{theobald2008spotsigs} \cite{hajishirzi2010adaptive} \cite{zhang2016effective}. Therefore, in order to compare the new method with other algorithms, the new approach is examined in this dataset.

\subsection{Tests}
\label{sec:4.2}
To test the new method, first, the desired reference text is generated using the genetic algorithm. The procedure explained in section 3.2 is used in order to generate the reference text. To find the best reference text, the cross-validation technique is used. The considered dataset was divided into two parts, one containing 80 percent and the other one 20 percent of the dataset's documents. The first part (80 percent) of the dataset is used in order to evaluate each reference text in a generation using fitness function. After completing the genetic algorithm process, the second part of the dataset (20 percent) is used in order to evaluate the final reference text. Due to the random nature of the genetic algorithm, this process is executed over ten times for each test, in order to average out the results and avoid very good or very bad random reference texts. Finally, the reference text with the highest fitness on the second part of the dataset (20 percent) is chosen as the winner. Then, the winner reference text and multi-reference cosine text similarity algorithm are used for detecting the DND documents in the other datasets. The mean absolute error of the algorithm is calculated using Eq. \ref{eq:1} for each test. For a DND text document detection, precision is the ratio of the number of correctly predicted DND text documents to the whole number of documents that are predicted as DND. On the other hand, recall is the ratio of the number of correctly predicted DND text documents to the whole number of DND text documents. F1-score is a weighted average of precision and recall. The equations to calculate the precision, recall, and the F1-score are as shown below (Eq. \ref{eq:2}, Eq. \ref{eq:3} and Eq. \ref{eq:4}):  

\begin{equation}
Precision = \frac{True\:Positives}{True\:Positives + False\:Positives}
\label{eq:2}
\end{equation}
\begin{equation}
Recall = \frac{True\:Positives}{True\:Positives + False\:Negatives}
\label{eq:3}
\end{equation}
\begin{equation}
F1\:score = 2 \times \frac{Precision \times Recall}{Precision+Recall}
\label{eq:4}
\end{equation}\\


Where true-positive is a document that is DND, and it is correctly predicted as DND by the system. False-positive is a document that initially is not a DND document, but the system has detected it as a DND document. Moreover, false-negative is a document that is originally a DND document, but the system incorrectly predicted it as a non-DND document. Fifty generations are used in order to find the best reference text because there is no significant change in the fitness of the best reference texts after 50 generations. In order to find a suitable population size, an experiment was performed. The genetic algorithm is executed to find the optimal reference text on the Citeseerx dataset with different population sizes. The test results are shown in Table \ref{tab:1}.

\begin{table}[h!]
\caption{Comparing performance of proposed method with different population sizes.}
\label{tab:1}
\begin{center}
\begin{tabular}{|l|l|l|l|}
\hline
Train Dataset & Population Size & Mean Absolute Error & Runtime of Each Iteration (seconds) \\
\hline
Citeseerx & 50 & 0.1083 & 9\\
Citeseerx & 100 & 0.1053 & 20\\
Citeseerx & 150 & 0.1041 & 30\\
\hline
\end{tabular}
\end{center}\vs{-4mm}
\end{table}

Table \ref{tab:1} shows that by increasing the population size, the final accuracy of the algorithm increases. However, this improvement is not worthy since the runtime of the genetic algorithm has a linear relation with the population size. So in order to have desired results and also reduce the runtime of the genetic algorithm, the population size of 100 chromosomes is used for the tests. \\
In Table \ref{tab:1}, the results of generating the reference text using each of datasets are presented. As mentioned above, 80 percent of documents of each dataset was used to evaluate the fitness of reference texts in the genetic algorithm. The remaining 20 percent of the dataset is used in order to evaluate the final reference text. In each of the following tests, reference texts with the length of 1000 3-grams were generated. The number of reference text partitions used in each test is 150. A reference text with 1000 3-grams length and 150 parts is used with the intention of the enhancing algorithm results by using these values. The cross-validation process is executed on the datasets. The final evaluation results are shown in Table \ref{tab:2}.

\begin{table}[h!]
\caption{Cross-validation evaluation results.}
\label{tab:2} 
\begin{center}
\begin{tabular}{|l|l|l|}
\hline
Train Dataset & Test Dataset & Mean Absolute Error\\
\hline
CiteseerX & CiteseerX & 0.1047 \\
TREC 05 & TREC 05 & 0.1246 \\
DMOZ & DMOZ & 0.1292 \\
20 Newsgroups & 20 Newsgroups & 0.1281 \\
OpenDNS public domains & OpenDNS public domains & 0.1070 \\
Enron & Enron & 0.1164 \\
\hline
\end{tabular}
\end{center}\vs{-4mm}
\end{table}

As shown in Table \ref{tab:2} the new approach has an average of 0.1183 difference in term of mean absolute error from the cosine text similarity algorithm. So it can be concluded that the multi-reference cosine text similarity algorithm is accurate and reliable. \\
In another test, the reference text is generated using the Citeseerx dataset as train dataset, and it is evaluated on the other datasets, to test the capability of the new method to be generalized to other datasets. As it was mentioned before, the algorithm achieves better results with a reference text with the length of 1000 3-grams and 150 parts. Therefore, this configuration can be used for this test too. The test results are shown in Table \ref{tab:3}.

\begin{table}[h!]
\caption{Training and evaluating over different datasets.}
\label{tab:3} 
\begin{center}
\begin{tabular}{|l|l|l|}
\hline
Reference text dataset & Evaluation Dataset & Mean Absolute Error\\
\hline
Citeseerx & TREC 05 & 0.1444 \\
Citeseerx & DMOZ & 0.1355 \\
Citeseerx & 20 Newsgroups & 0.1012 \\
Citeseerx & OpenDNS public domains & 0.1355 \\
Citeseerx & Enron & 0.1389 \\
\hline
\end{tabular}
\end{center}\vs{-4mm}
\end{table}

As shown in Table \ref{tab:3}, the test results on the other datasets are a little bit worse but generally are promising.
In order to understand the effects of reference text length and the number of partitions on the accuracy of the final results, reference texts with different lengths and also using different number of partitions are generated. Citeseerx dataset is used for these tests. The results are shown in Table \ref{tab:4}.

\begin{table}[h!]
\caption{Test results using different reference text lengths and different number of partitions.}
\label{tab:4}
\begin{center}
\begin{tabular}{|l|l|l|}
\hline
Reference text length & Number of partitions & Mean absolute error \\
\hline
450 3-grams & 100 & 0.1226 \\
450 3-grams & 150 & 0.1240 \\
450 3-grams & 300 & 0.1281 \\
600 3-grams & 100 & 0.1190 \\
600 3-grams & 150 & 0.1186 \\
600 3-grams & 300 & 0.1197 \\
1000 3-grams & 100 & 0.1193 \\
1000 3-grams & 150 & 0.1047 \\
1000 3-grams & 300 & 0.1135 \\
\hline
\end{tabular}
\end{center}\vs{-4mm}
\end{table}

Table \ref{tab:4} shows that, in general, increasing the length of reference text increases the accuracy of multi-reference cosine text similarity algorithm. This was discussed in section \ref{sec:3.6.1}. Furthermore, the results show that the relation between the number of partitions and the final accuracy is somehow ambiguous. Nevertheless, it seems that for different lengths of the reference text, dividing the reference text into 150 parts leads to better results. Table \ref{tab:4} shows that very long or very short reference partitions, decrease the algorithm accuracy. \\
The new approach is tested on the Gold Set of Near-Duplicate News Articles in order to compare the new algorithm with the other recent approaches. The precision and recall of other approaches are collected from a paper from Zhang et al. named "Effective and Fast Near Duplicate Detection via Signature-Based Compression Metrics" \cite{zhang2016effective}. For this test, a reference text is used, that is generated using Citeseerx dataset, with the length of 1000 and 150 parts. The precision, recall, and the F1-score of the proposed algorithm and other algorithm are shown in Table \ref{tab:5}.

\begin{table}[h!]
\caption{Comparing proposed method with other approaches}
\label{tab:5} 
\begin{center}
\begin{tabular}{|l|l|l|l|l|}
\hline
Algorithm & Precision & Recall & F1-score & Runtime (ms) \\
\hline
SigNCD & 0.95 & 0.89 & 0.92 & 7838\\
SpotSigNCD & 0.97 & 0.87 & 0.92 & 10853\\
Proposed method & 0.87 & 0.98 & 0.92 & 10950 \\
NCD & 0.90 & 0.78 & 0.83 & 53829\\
SpotSigs & 0.90 & 0.77 & 0.83 & 12824\\
Simhash & 0.85 & 0.45 & 0.59 & 13010\\
\hline
\end{tabular}
\end{center}\vs{-4mm}
\end{table}

As presented in Table \ref{tab:5}, multi-reference cosine achieved an F1-score equivalent to the state of the art algorithms. In addition to that, the proposed algorithm obtained the highest recall between the algorithms presented in Table \ref{tab:5}, which gives it special confident in detecting the highest number of DND documents in a target document collection. The runtime measured for the proposed algorithm is the time required for its comparisons and does not include the training section of the proposed algorithm. Considering the runtime, we cannot precisely compare the new algorithm's runtime with other algorithms because the runtime of the new algorithm is computed on different hardware than what is used in order to compute the runtime for other algorithms. However, the runtime of the new algorithm is approximately close to the best algorithms, regarding the little hardware specification differences.

\section{Conclusions and Future works}
\label{sec:5}
In this paper a new method to generate better reference texts to be used in multi-reference cosine text similarity algorithm is proposed. Multi-reference cosine algorithm shows good results in large collections of text documents. By using better reference texts, its accuracy will be improved significantly. The new approach shows reliable and promising results. The new algorithm can achieve close to cosine text similarity algorithm's accuracy, but by consuming less computational power. Furthermore, the algorithm achieves a better recall in comparison with other recent methods. The novel approach of multi-reference cosine algorithm toward measuring the similarity between two documents using a reference text makes the achievement of such high recall feasible. As stated in section \ref{sec:3.2}, the reference texts are initiated and mutated using 3-grams with high Tf-idf score. Consequently, the final reference text consists of several high Tf-idf score 3-grams that are highly beneficial in term of detecting the similarity between two documents. This quality of this algorithm makes it more sensitive to the similarity between two documents and therefore delivers a high recall. Thus as the new method has better recall than other state of the art methods, it is more suitable for practical application because the system could assure the detection of all DND documents. \\
As future works, this algorithm can be adapted to map-reduce programming framework to be used in big data distributed frameworks such as Hadoop. Finding new ways to combine multiple reference texts that are generated using different datasets in order to achieve better general accuracy for diverse applications, is another interesting future work. Extending the method to be able to work with multiple languages could be considered as another good study to focus on.

\bibliographystyle{IEEEtran}
\bibliography{mybibfile.bib}

\begin{thebibliography}{10}
\providecommand{\url}[1]{#1}
\csname url@samestyle\endcsname
\providecommand{\newblock}{\relax}
\providecommand{\bibinfo}[2]{#2}
\providecommand{\BIBentrySTDinterwordspacing}{\spaceskip=0pt\relax}
\providecommand{\BIBentryALTinterwordstretchfactor}{4}
\providecommand{\BIBentryALTinterwordspacing}{\spaceskip=\fontdimen2\font plus
\BIBentryALTinterwordstretchfactor\fontdimen3\font minus
  \fontdimen4\font\relax}
\providecommand{\BIBforeignlanguage}[2]{{%
\expandafter\ifx\csname l@#1\endcsname\relax
\typeout{** WARNING: IEEEtran.bst: No hyphenation pattern has been}%
\typeout{** loaded for the language `#1'. Using the pattern for}%
\typeout{** the default language instead.}%
\else
\language=\csname l@#1\endcsname
\fi
#2}}
\providecommand{\BIBdecl}{\relax}
\BIBdecl

\bibitem{broder1997syntactic}
A.~Z. Broder, S.~C. Glassman, M.~S. Manasse, and G.~Zweig, ``Syntactic
  clustering of the web,'' \emph{Computer Networks and ISDN Systems}, vol.~29,
  no. 8-13, pp. 1157--1166, 1997.

\bibitem{fetterly2003evolution}
D.~Fetterly, M.~Manasse, and M.~Najork, ``On the evolution of clusters of
  near-duplicate web pages,'' \emph{Journal of Web Engineering}, vol.~2, no.~4,
  pp. 228--246, 2003.

\bibitem{henzinger2006finding}
M.~Henzinger, ``Finding near-duplicate web pages: a large-scale evaluation of
  algorithms,'' in \emph{Proceedings of the 29th annual international ACM SIGIR
  conference on Research and development in information retrieval}.\hskip 1em
  plus 0.5em minus 0.4em\relax ACM, 2006, pp. 284--291.

\bibitem{manku2007detecting}
G.~S. Manku, A.~Jain, and A.~Das~Sarma, ``Detecting near-duplicates for web
  crawling,'' in \emph{Proceedings of the 16th international conference on
  World Wide Web}.\hskip 1em plus 0.5em minus 0.4em\relax ACM, 2007, pp.
  141--150.

\bibitem{xiao2011efficient}
C.~Xiao, W.~Wang, X.~Lin, J.~X. Yu, and G.~Wang, ``Efficient similarity joins
  for near-duplicate detection,'' \emph{ACM Transactions on Database Systems
  (TODS)}, vol.~36, no.~3, p.~15, 2011.

\bibitem{conrad2003online}
J.~G. Conrad, X.~S. Guo, and C.~P. Schriber, ``Online duplicate document
  detection: signature reliability in a dynamic retrieval environment,'' in
  \emph{Proceedings of the twelfth international conference on Information and
  knowledge management}.\hskip 1em plus 0.5em minus 0.4em\relax ACM, 2003, pp.
  443--452.

\bibitem{theobald2008spotsigs}
M.~Theobald, J.~Siddharth, and A.~Paepcke, ``Spotsigs: robust and efficient
  near duplicate detection in large web collections,'' in \emph{Proceedings of
  the 31st annual international ACM SIGIR conference on Research and
  development in information retrieval}.\hskip 1em plus 0.5em minus 0.4em\relax
  ACM, 2008, pp. 563--570.

\bibitem{multireferencecosine}
H.~Mohammadi and A.~Nikoukaran, ``Multi-reference cosine: A new approach to
  text similarity measurement in large collections,'' \emph{arXiv preprint
  arXiv:1810.03099}, 2018.

\bibitem{ntoulas2004s}
A.~Ntoulas, J.~Cho, and C.~Olston, ``What's new on the web?: the evolution of
  the web from a search engine perspective,'' in \emph{Proceedings of the 13th
  international conference on World Wide Web}.\hskip 1em plus 0.5em minus
  0.4em\relax ACM, 2004, pp. 1--12.

\bibitem{croft2010search}
W.~B. Croft, D.~Metzler, and T.~Strohman, \emph{Search engines: Information
  retrieval in practice}.\hskip 1em plus 0.5em minus 0.4em\relax Addison-Wesley
  Reading, 2010, vol. 283.

\bibitem{koppula2010learning}
H.~S. Koppula, K.~P. Leela, A.~Agarwal, K.~P. Chitrapura, S.~Garg, and
  A.~Sasturkar, ``Learning url patterns for webpage de-duplication,'' in
  \emph{Proceedings of the third ACM international conference on Web search and
  data mining}.\hskip 1em plus 0.5em minus 0.4em\relax ACM, 2010, pp. 381--390.

\bibitem{bar2009not}
Z.~Bar-Yossef, I.~Keidar, and U.~Schonfeld, ``Do not crawl in the dust:
  different urls with similar text,'' \emph{ACM Transactions on the Web
  (TWEB)}, vol.~3, no.~1, p.~3, 2009.

\bibitem{dasgupta2008duping}
A.~Dasgupta, R.~Kumar, and A.~Sasturkar, ``De-duping urls via rewrite rules,''
  in \emph{Proceedings of the 14th ACM SIGKDD international conference on
  Knowledge discovery and data mining}.\hskip 1em plus 0.5em minus 0.4em\relax
  ACM, 2008, pp. 186--194.

\bibitem{shivakumar1995scam}
N.~Shivakumar and H.~Garcia-Molina, ``Scam: A copy detection mechanism for
  digital documents,'' 1995.

\bibitem{manber1994finding}
U.~Manber \emph{et~al.}, ``Finding similar files in a large file system.'' in
  \emph{Usenix Winter}, vol.~94, 1994, pp. 1--10.

\bibitem{heintze1996scalable}
N.~Heintze \emph{et~al.}, ``Scalable document fingerprinting,'' in \emph{1996
  USENIX workshop on electronic commerce}, vol.~3, no.~1, 1996.

\bibitem{brin1995copy}
S.~Brin, J.~Davis, and H.~Garcia-Molina, ``Copy detection mechanisms for
  digital documents,'' in \emph{ACM SIGMOD Record}, vol.~24, no.~2.\hskip 1em
  plus 0.5em minus 0.4em\relax Acm, 1995, pp. 398--409.

\bibitem{indyk1998approximate}
P.~Indyk and R.~Motwani, ``Approximate nearest neighbors: towards removing the
  curse of dimensionality,'' in \emph{Proceedings of the thirtieth annual ACM
  symposium on Theory of computing}.\hskip 1em plus 0.5em minus 0.4em\relax
  ACM, 1998, pp. 604--613.

\bibitem{schleimer2003winnowing}
S.~Schleimer, D.~S. Wilkerson, and A.~Aiken, ``Winnowing: local algorithms for
  document fingerprinting,'' in \emph{Proceedings of the 2003 ACM SIGMOD
  international conference on Management of data}.\hskip 1em plus 0.5em minus
  0.4em\relax ACM, 2003, pp. 76--85.

\bibitem{chowdhury2002collection}
A.~Chowdhury, O.~Frieder, D.~Grossman, and M.~C. McCabe, ``Collection
  statistics for fast duplicate document detection,'' \emph{ACM Transactions on
  Information Systems (TOIS)}, vol.~20, no.~2, pp. 171--191, 2002.

\bibitem{sarawagi2004efficient}
S.~Sarawagi and A.~Kirpal, ``Efficient set joins on similarity predicates,'' in
  \emph{Proceedings of the 2004 ACM SIGMOD international conference on
  Management of data}.\hskip 1em plus 0.5em minus 0.4em\relax ACM, 2004, pp.
  743--754.

\bibitem{charikar2002similarity}
M.~S. Charikar, ``Similarity estimation techniques from rounding algorithms,''
  in \emph{Proceedings of the thiry-fourth annual ACM symposium on Theory of
  computing}.\hskip 1em plus 0.5em minus 0.4em\relax ACM, 2002, pp. 380--388.

\bibitem{pamulaparty2013novel}
L.~Pamulaparty and C.~G. Rao, ``A novel approach to perform document clustering
  using effectiveness and efficiency of simhash,'' \emph{International Journal
  of Engineering and Advanced Technology}, vol.~2, no.~3, pp. 312--315, 2013.

\bibitem{hajishirzi2010adaptive}
H.~Hajishirzi, W.-t. Yih, and A.~Kolcz, ``Adaptive near-duplicate detection via
  similarity learning,'' in \emph{Proceedings of the 33rd international ACM
  SIGIR conference on Research and development in information retrieval}.\hskip
  1em plus 0.5em minus 0.4em\relax ACM, 2010, pp. 419--426.

\bibitem{lin2009brute}
J.~Lin, ``Brute force and indexed approaches to pairwise document similarity
  comparisons with mapreduce,'' in \emph{Proceedings of the 32nd international
  ACM SIGIR conference on Research and development in information
  retrieval}.\hskip 1em plus 0.5em minus 0.4em\relax ACM, 2009, pp. 155--162.

\bibitem{vernica2010efficient}
R.~Vernica, M.~J. Carey, and C.~Li, ``Efficient parallel set-similarity joins
  using mapreduce,'' in \emph{Proceedings of the 2010 ACM SIGMOD International
  Conference on Management of data}.\hskip 1em plus 0.5em minus 0.4em\relax
  ACM, 2010, pp. 495--506.

\bibitem{pamulaparty2014near}
L.~Pamulaparty, C.~G. Rao, and M.~S. Rao, ``A near-duplicate detection
  algorithm to facilitate document clustering,'' \emph{International Journal of
  Data Mining \& Knowledge Management Process}, vol.~4, no.~6, p.~39, 2014.

\bibitem{varol2015detecting}
C.~Varol and S.~Hari, ``Detecting near-duplicate text documents with a hybrid
  approach,'' \emph{Journal of Information Science}, vol.~41, no.~4, pp.
  405--414, 2015.

\bibitem{zhang2016effective}
X.~Zhang, Y.~Yao, Y.~Ji, and B.~Fang, ``Effective and fast near duplicate
  detection via signature-based compression metrics,'' \emph{Mathematical
  Problems in Engineering}, vol. 2016, 2016.

\bibitem{cilibrasi2005clustering}
R.~Cilibrasi and P.~M. Vit{\'a}nyi, ``Clustering by compression,'' \emph{IEEE
  Transactions on Information theory}, vol.~51, no.~4, pp. 1523--1545, 2005.

\bibitem{li2008p}
M.~Li and M.~Paul, ``P. vit anyi, an introduction to kolmogorov complexity and
  its applications,'' 2008.

\bibitem{liu2014new}
H.~Liu, Z.~Hu, A.~Mian, H.~Tian, and X.~Zhu, ``A new user similarity model to
  improve the accuracy of collaborative filtering,'' \emph{Knowledge-Based
  Systems}, vol.~56, pp. 156--166, 2014.

\bibitem{ide2009american}
N.~Ide, ``The american national corpus: Then, now, and tomorrow,'' in
  \emph{Selected Proceedings of the 2008 HC-SNet Workshop on Designing an
  Australian National Corpus}, 2009, pp. 108--113.

\bibitem{Citeseerx}
``Citeseerx data. (2017) [data file],''
  \url{http://csxstatic.ist.psu.edu/about/data}, accessed: May 5 2017.

\bibitem{TREC}
``Trec 2005 public spam corpus. (2005) [data file],''
  \url{https://plg.uwaterloo.ca/~gvcormac/treccorpus/}, accessed: May 5 2017.

\bibitem{DMOZ}
``Dmoz, (2016) [data file],''
  \url{https://dataverse.harvard.edu/dataset.xhtml?persistentId=doi:10.7910/DVN/OMV93V},
  accessed: May 5 2017.

\bibitem{Newsgroups}
K.~Lang, ``Ken lang, (1995), 20 newsgroups [data file],''
  \url{http://qwone.com/~jason/20Newsgroups/}, accessed: October 26 2016.

\bibitem{lang1995newsweeder}
------, ``Newsweeder: Learning to filter netnews,'' in \emph{Machine Learning
  Proceedings 1995}.\hskip 1em plus 0.5em minus 0.4em\relax Elsevier, 1995, pp.
  331--339.

\bibitem{OpenDNS}
``Opendns top domains list, (2014) [data file],''
  \url{https://github.com/opendns/public-domain-lists}, accessed: May 5 2017.

\bibitem{Enron}
``Enron, (2015) [data file],'' \url{https://www.cs.cmu.edu/~./enron/},
  accessed: January 7 2017.

\bibitem{GoldSet}
``Gold set of near-duplicate news articles, (2008) [data file],''
  \url{http://adrem.ua.ac.be/~tmartin/}, accessed: January 7 2017.

\bibitem{cho2006stanford}
J.~Cho, H.~Garcia-Molina, T.~Haveliwala, W.~Lam, A.~Paepcke, S.~Raghavan, and
  G.~Wesley, ``Stanford webbase components and appfcations,'' \emph{ACM
  Transactions on Internet Technology (TOIT)}, vol.~6, no.~2, pp. 153--186,
  2006.

\end{thebibliography}

\end{document}